\documentstyle[preprint,aps]{revtex}
\draft
\begin{document}
\title{Compressible phase of a double-layer electron system with total
Landau-level filling factor one-half}
\author{N.~E.~Bonesteel}
\address{
National High Magnetic Field Laboratory, Florida State University,
P.O. Box 4005, Tallahassee, FL 32306-4005}
\maketitle
\begin{abstract}
\noindent
Following recent work of Halperin, Lee, and Read, and Kalmeyer and
Zhang, a double-layer electron system with total Landau-level filling
factor $\nu=1/2$ is mapped onto an equivalent system of fermions in
zero average magnetic field interacting via a Chern-Simons gauge
field. Within the random-phase approximation a new, low-lying,
diffusive mode, not present in the $\nu=1/2$ single-layer system, is
found.  This mode leads to more singular low-energy scattering than
appears in the single layer system, and to an attractive pairing
interaction between fermions in different layers which grows stronger
as the layer spacing is decreased.  The possible connection between
this pairing interaction and the experimentally observed fractional
quantum Hall effect in double-layer systems is discussed.
\end{abstract}
\pacs{}

Recently, the fractional quantum Hall effect (FQHE) has been seen in
double-layer electron systems with total Landau-level filling factor
$\nu = 1/2$ \cite{eisenstein,suen}.  This observation supports the
long-held belief that incompressible, even-denominator quantum Hall
states can exist when there are two species of fermions
\cite{halperin-ss,yoshioka,he,haldane}.  However, in these systems,
as the ratio of the layer spacing, $d$, to the magnetic length, $l_0$
($\equiv (\hbar c/eB_0)^{1/2}$, where $B_0$ is the applied magnetic
field), is increased, the FQHE becomes weaker and eventually
disappears \cite{eisenstein}, indicating that the state has become
compressible.  The $\nu=1/2$ single-layer system, which is also
compressible \cite{1layer}, has been described by Halperin, Lee, and
Read \cite{halperin}, and Kalmeyer and Zhang \cite{kalmeyer}, in terms
of a `Fermi liquid' of electrons bound to an even number of flux
quanta.  It is the purpose of this paper to develop a similar
description for the compressible phase of the $\nu=1/2$ double-layer
system, and to point out some new features of this description which
are not present in the single layer case.

Specifically, we consider a double-layer system in a transverse
magnetic field, $B_0 = 4hc n/e$, where $n$ is the carrier density per
layer, (for the remainder of this paper $\hbar = c = 1$, so that, {\it
e.g.}, $B_0 = 8\pi n/e$).  The total Landau-level filling factor is
then $\nu=1/2$, and the filling factor in each layer is $\nu=1/4$.
Electron spins are assumed to be fully polarized, and tunneling
between layers is ignored.

In two dimensions, it is possible to continuously change the
statistics of identical particles by attaching infinitely thin flux
tubes containing fictitious flux to them \cite{wilczek}.  In this
paper we will refer to this fictitious flux as Chern-Simons flux, and
to the transformed particles as pseudo-particles.  In double-layer
systems it is useful to introduce {\it two} types of Chern-Simons
flux.  The relationship between physical electrons and
pseudo-particles can then be expressed mathematically as a `singular
gauge transformation' of the form
\begin{eqnarray}
\psi^{\phantom{\dagger}}_{1}({\bf r}) &=&
\psi^{\phantom{\dagger}}_{e,1}({\bf r})\exp
\biggl( i\phi_1\int d^2r^\prime \arg({\bf r}-{\bf r}^\prime)
\rho_1({\bf r}^\prime)
+i\phi_2\int d^2r^\prime \arg({\bf r}-{\bf r}^\prime)
\rho_2({\bf r}^\prime)
\biggr),\nonumber\\
\psi^{\phantom{\dagger}}_{2}({\bf r}) &=&
\psi^{\phantom{\dagger}}_{e,2}({\bf r})\exp
\biggl(
i\phi_2\int d^2r^\prime \arg({\bf r}-{\bf r}^\prime)
\rho_1({\bf r}^\prime)
+i\phi_1\int d^2r^\prime \arg({\bf r}-{\bf r}^\prime)
\rho_2({\bf r}^\prime)
\biggr),
\end{eqnarray}
where $\psi_{e,s}(r)$ and $\psi_s(r)$ are, respectively, the physical
electron and pseudo-particle annihilation operators in layer $s$,
$\rho_s(r)$ is the density operator in layer $s$, and $\arg({\bf
r}-{\bf r}^\prime)$ is the angle made by the vector ${\bf r}-{\bf
r}^\prime$ and the $x$-direction, (throughout this paper all spatial
vectors are projected into the $xy$ plane).  A pseudo-particle in a
given layer then sees $\phi_1$ flux quanta attached to particles in
that same layer, and $\phi_2$ flux quanta attached to particles in the
next layer.  In the absence of interlayer tunneling, the relative
statistics of particles in different layers is irrelevant, and the
statistics of the pseudo-particles depends only on $\phi_1$.  In
particular, the pseudo-particles are bosons if $\phi_1$ is odd
\cite{zhang}, and fermions if $\phi_1$ is even \cite{jain}.

For the $\nu=1/2$ double-layer system there are several interesting
choices for $(\phi_1,\phi_2)$.  For example, if $(\phi_1,\phi_2) =
(3,1)$ the pseudo-particles obey Bose statistics.  When the
Chern-Simons flux attached to these particles is smeared out according
to the standard mean-field prescription \cite{semions}, each
pseudo-boson sees an effective average field ${\overline B} = B_0 -
2\pi(\phi_1+\phi_2)n/e = 0$ and the applied magnetic field is
cancelled exactly. The pseudo-bosons then condense and, just as for
the single-layer $\nu=1/3$ FQHE, the resulting condensed state can be
shown to be incompressible \cite{zhang}.  The wave function
corresponding to this condensed state is, presumably, the so-called
331 wave function, a generalization of Laughlin's wave function for
double-layer systems
\cite{halperin-ss,yoshioka,he}.  Exact diagonalization studies have
shown that the 331 wave function has a significant overlap with the
exact ground state for finite size systems only when $d/l_0 \alt 4$
\cite{yoshioka,he}.  This is consistent with the fact that the (3,1)
scheme becomes untenable in the $d\rightarrow\infty$ limit, because
particles in one layer continue to see flux attached to particles in
the second layer.  To avoid this problem, but retain the mean-field
cancellation of the physical field, there is a unique choice:
$(\phi_1,\phi_2) = (4,0)$.  The pseudo-particles are then fermions,
and, in the $d\rightarrow\infty$ limit, the layers decouple naturally,
so that each layer is described by a compressible `Fermi liquid' of
pseudo-fermions \cite{halperin,kalmeyer}.

To study the (4,0) mean field theory plus Gaussian fluctuations for
finite $d$, it is convenient to formulate the problem in terms of a
finite-temperature, Euclidean-time functional integral.  For
$(\phi_1,\phi_2) = (\phi,0)$, where $\phi=4$ when $\nu=1/2$, the
Lagrangian density describing the pseudo-fermions is $L({\bf r},\tau)
= L_0({\bf r},\tau) + L_1({\bf r},\tau)$, where the first part of the
Lagrangian is
\begin{eqnarray}
L_0({\bf r},\tau) = \sum_{s=1,2}\Biggl(&&\psi_s^*({\bf
r},\tau)(\partial_t-ia^{(s)}_0({\bf r},\tau))\psi_s({\bf
r},\tau)\nonumber\\ +&&{1\over{2m_b}}\psi_s^*({\bf r},\tau)
(-i{\bbox{\nabla}} + {\bf a}^{(s)}({\bf r},\tau)- e{\bf A}_0({\bf
r}))^2 \psi_s({\bf r},\tau) \Biggr).\label{l0}
\end{eqnarray}
Here $\psi_s$, and $(a^{(s)}_0,{\bf a}^{(s)})$ are, respectively, the
pseudo-fermion field and a Chern-Simons gauge field in layer $s$,
${\bf A}_0({\bf r}) = ({\hat {\bf z}}\times{\bf r})B_0/2$ is the
physical vector potential describing the applied magnetic field, and
$m_b$ is the band mass of the electrons. The second part of the
Lagrangian,
\begin{eqnarray}
L_1({\bf r},\tau) &=& \sum_{\scriptstyle s=1,2
\atop{\scriptstyle
{\phantom{s^\prime=1,2}}}} {-i\over{2\pi\phi}}a^{(s)}_0({\bf
r},\tau){\hat{\bf z}}\cdot( {\bbox{\nabla}}\times{\bf a}^{(s)}({\bf
r},\tau))\nonumber\\
&+&\sum_{s,s^\prime =1,2} {1\over{2(2\pi\phi)^2}}
\int d^2{r^\prime} ({\bbox{\nabla}}\times{\bf
a}^{(s)}({\bf r},\tau)) V_{s,s^\prime}({\bf r}-{\bf r}^\prime)
({\bbox{\nabla}}\times{\bf a}^{(s)}({\bf r}^\prime,\tau)),\label{l1}
\end{eqnarray}
depends only on the Chern-Simons gauge fields.  We work in the
Coulomb gauge, ${\bbox{\nabla}}\cdot{\bf a}^{(s)}({\bf r},\tau) = 0$,
where the first term in (\ref{l1}) is the Chern-Simons term.
Integrating out the time components of the Chern-Simons gauge fields
then enforces the constraint \cite{zhang}
\begin{equation}
2\pi\phi \psi_s^*({\bf
r},\tau)\psi_s({\bf r},\tau) = {\hat{\bf z}}\cdot
({\bbox{\nabla}}\times{\bf a}^{(s)}({\bf r},\tau))\label{constraint}
\end{equation}
which describes attaching $\phi$ fictitious $a^{(s)}$ flux quanta to
each pseudo-fermion in layer $s$.  The second term in (\ref{l1}) is
the interlayer and intralayer Coulomb repulsion, where, following
Ref.~\cite{halperin}, the constraint (\ref{constraint}) has been used
to rewrite this term purely in terms of the Chern-Simons gauge fields.
Finally, the Coulomb repulsion itself is
\begin{eqnarray}
V_{ss^\prime}({\bf r}) = {2\pi
e^2\over{\varepsilon\sqrt{r^2+d^2(1-\delta_{s,s^\prime})}}}
\end{eqnarray}
where $\varepsilon$ is the dielectric constant.

At the mean-field level the Chern-Simons gauge fields take their
average values, $\langle {\bf a}^{(s)}({\bf r},\tau)\rangle = 2\pi\phi
\langle \rho^{(s)}({\bf r})\rangle ({\hat{\bf z}}\times{\bf r}) =
2\pi\phi n({\hat{\bf z}}\times{\bf r})$, and, for $\nu=1/2$ and
$\phi=4$, the applied magnetic field is cancelled exactly.  The
pseudo-fermions in each layer then form Fermi liquids with Fermi wave
vector $k_f = (4\pi n)^{1/2} = (1/l_0)(2/\phi)^{1/2}$.  Gaussian
fluctuations about this mean field state can be studied by integrating
out the pseudo-fermion fields in (\ref{l0}), exponentiating the
resulting determinant, and expanding to one loop order in the
Chern-Simons gauge fields, a procedure which is equivalent to the
random-phase approximation developed for the anyon gas \cite{semions}.
The result is an effective action for the Chern-Simons gauge fields,
\begin{equation}
S_{\rm eff}[a] = {1\over{2}}\sum_n \int
{d^2q\over{(2\pi)^2}}\sum_{\scriptstyle
s,s^\prime=1,2\atop{\scriptstyle
\mu,\nu=0,1}} a^{(s)}_\mu({\bf q},i\omega_n)
 D^{-1}_{s,\mu;s^\prime,\nu}({\bf q},i\omega_n)
a^{(s^\prime)}_\nu(-{\bf q},-i\omega_n),
\end{equation}
where $a^{(s)}_1({\bf q},\omega) = {\hat{\bf z}}\cdot({\bf{\hat
q}}\times ({\bf a}^{(s)}({\bf q},\omega)-\langle {\bf a}^{(s)}({\bf
q},\omega)\rangle))$ is the fluctuation in the transverse gauge field,
and the matrix $D_{s,\mu;s^\prime,\nu}({\bf q},\omega)$ is the
effective propagator for the Chern-Simons gauge fields, the inverse of
which is
\begin{equation}
D^{-1}_{s,\mu;s^\prime,\nu}({\bf q},i\omega_n)
=
\left(
\begin{array}{cccc}
\Pi^0_{00} & {{\displaystyle iq}\over{\displaystyle 2\pi\phi}} & 0 & 0
\\-{{\displaystyle iq}\over{\displaystyle 2\pi\phi}}
& \Pi^0_{11} - {\displaystyle{q^2 V_{11}(q)}
\over{\displaystyle (2\pi\phi)^2}}& 0 & -{{\displaystyle
q^2V_{12}(q)}\over{\displaystyle{(2\pi\phi)^2}}}
\\ 0 & 0 & \Pi^0_{00} & {{\displaystyle iq}
\over{\displaystyle 2\pi\phi}} \\ 0 &
-{\displaystyle{q^2V_{12}(q)}\over{\displaystyle{(2\pi\phi)^2}}}
& -{{\displaystyle iq}\over{\displaystyle 2\pi\phi}} & \Pi^0_{11}
- {\displaystyle{q^2 V_{11}(q)}
\over{\displaystyle{(2\pi\phi)^2}}}\\
\end{array}
\right)\label{invprop}
\end{equation}
This matrix is labeled according to the scheme $(s,\mu) =
[(1,0),(1,1),(2,0),(2,1)]$, where $s$ is the layer index, and $\mu
=0,1$ labels the time and transverse component of the Chern-Simons
gauge fields.  Finally, the noninteracting density and
transverse-current polarization functions appearing in
(\ref{invprop}), are, respectively,
\begin{equation}
\Pi^0_{00} = \int {d^2k\over{(2\pi)^2}} {f(\epsilon_{{\bf k}+{\bf
q}})-f(\epsilon_{\bf k})\over{i\omega_n-\epsilon_{{\bf k}+{\bf
q}}+\epsilon_{\bf k}}},\label{longresp}
\end{equation}
and
\begin{equation}
\Pi^0_{11} = \int {d^2k\over{(2\pi)^2}}\left({\hat{\bf q}}\times{\bf
k}\over{m_b}\right)^2 {f(\epsilon_{{\bf k}+{\bf q}})-f(\epsilon_{\bf
k})\over{i\omega_n-\epsilon_{{\bf k}+{\bf q}}+\epsilon_{\bf k}}}-{
n\over{m_b}},\label{transresp}
\end{equation}
where $\epsilon_{\bf k} = k^2/2m_b - \mu_f$, $\mu_f$ is the chemical
potential, and $f(\epsilon_{\bf k})$ is the Fermi function.

The collective modes of the system correspond to poles in the gauge
field propagator and can be found by solving the equation $\det
D^{-1}(q,\omega) = 0$.  In the limit $\omega \gg k_f q/m_b$ the
analytically continued polarization functions (\ref{longresp}) and
(\ref{transresp}) are $\Pi^0_{00} \simeq -(n/m_b)(q^2/\omega^2)$ and
$\Pi^0_{11} \simeq - n/m_b$.  These expressions can be used to find
two propagating modes with dispersion relations
\begin{equation}
\omega^{(1)}(q) \simeq \omega_c + {e^2\over{2\varepsilon\phi}}q,
\label{kohn1}
\end{equation}
and
\begin{equation}
\omega^{(2)}(q) \simeq \omega_c + {e^2 d\over{2\varepsilon\phi}}q^2,
\label{kohn2}
\end{equation}
where $\omega_c=2\pi\phi n/m_b$ is the cyclotron frequency.  In the
limit $q \ll k_f$, $\omega \ll k_f q/m_b$, the polarization functions
are $\Pi^0_{00} \simeq m_b/2\pi$ and $\Pi^0_{11} \simeq -\chi_{\rm d}
q^2 + i k_f\omega/4\pi q$, where $\chi_{d} = (12\pi m_b)^{-1}$ is the
Landau diamagnetic susceptibility for noninteracting electrons.  Again
there are two modes, this time diffusive, with dispersion relations
\begin{equation}
\omega^{(3)}(q) \simeq i
{4 e^2\over{\varepsilon k_f\phi^2}}q^2,\label{diff1}
\end{equation}
and
\begin{equation}
\omega^{(4)}(q) \simeq i {{4\pi\tilde\chi}\over{k_f}}q^3,\label{diff2}
\end{equation}
where
\begin{equation}
\tilde\chi = {(1+e^2 m_b d)\over{2\pi\phi^2 m_b}}+{1\over{12\pi m_b}}.
\label{chi}
\end{equation}
It is interesting to compare these modes with those which appear in
the Fermi-liquid description of the $\nu=1/2$ single-layer system
\cite{halperin}.  When the long-range Coulomb repulsion between
electrons is properly taken into account, the collective modes of the
single-layer system have the same dispersions as modes 1 and 3, while
for short-ranged interactions they have the same dispersions as modes
2 and 4 \cite{halperin}.  This correspondence is not surprising,
because in the double-layer system modes 1 and 3 involve density
fluctuations which are in phase in the two layers, while modes 2 and 4
involve density fluctuations which are out of phase, and hence are
unaffected by the long-range nature of the Coulomb repulsion.  We note
in passing that the high-energy propagating modes 1 and 2 are the
in-phase, and out-of-phase magneto-plasmons, or Kohn's modes, and
(\ref{kohn1}) is consistent with Kohn's theorem \cite{kohn}, which
requires that the energy of mode 1 go to the unrenormalized cyclotron
frequency, $\omega_c$, as $q\rightarrow 0$.  Within the random-phase
approximation, the energy of mode 2 also goes to $\omega_c$ as
$q\rightarrow 0$, and while this is not required by Kohn's theorem,
which follows from translational symmetry and thus only applies to the
in-phase mode, it does reflect the `decoupled' nature of the (4,0)
state.

As in the single-layer system, the diffusive modes 3 and 4 are the
most important source of low-energy quasiparticle scattering.  To
study these modes it is useful to define the symmetric and
antisymmetric transverse Chern-Simons gauge fields: $a^{(\pm)}_1 =
(a_1^{(1)} \pm a_1^{(2)}) /\sqrt{2}$.  The retarded propagators for
these fields are found by performing the matrix inversion
(\ref{invprop}) and analytically continuing to the real frequency
axis.  In the $q \ll k_f$, $\omega
\ll k_f q/m_b$ limit, the resulting propagators are
\begin{equation}
D_{11;+}({\bf q},\omega) \simeq{1\over{2}}
\left[{e^2\over{\pi\phi^2}}q -i{k_f\omega\over{4\pi q}} \right]^{-1},
\end{equation}
and
\begin{equation}
D_{11;-}({\bf q},\omega) \simeq{1\over{2}}
\left[\tilde\chi q^2 -i{k_f\omega\over{4\pi q}}\right]^{-1}
\end{equation}
for the symmetric and antisymmetric transverse Chern-Simons gauge
fields, respectively.  At low frequencies and long wavelengths, the
effective interaction between pseudo-fermions is dominated by
$D_{11;-}$, (all other components of $D$ are less singular for small
$q$), and has the form
\begin{equation}
V^{\rm eff}_{ss^\prime}({\bf k},{\bf k}^\prime;{\bf q},\omega) \simeq
\left(
\begin{array}{cc}
-1  & \phantom{-}1 \\
\phantom{-}1 & -1 \\
\end{array}
\right)
{({\bf k}\times {\hat{\bf q}})\cdot({\bf k^\prime}\times {\hat{\bf
q}})
\over{m_b^2}}
{1\over{2}}\left[\tilde\chi q^2-i{ k_f\omega\over{4\pi
q}}\right]^{-1}\label{veff3}
\end{equation}
where the matrix is a layer matrix.  It follows from (\ref{chi}) that
this interaction grows stronger with decreasing $d$.  Precisely such a
singular current-current interaction appears in the $\nu=1/2$
single-layer system for the physically unrealistic case of
short-ranged electron-electron interactions
\cite{halperin}. In the double-layer system, not only does this
interaction appear even when the long-range Coulomb repulsion is
included, but, when ${\bf k}^\prime = -{\bf k}$, {\it i.e.}, in the
Cooper channel, {\it the effective interaction between pseudo-fermions
in different layers is attractive.}

Physically, this attractive pairing interaction appears because
$a^{(-)}_1$ couples to pseudo-fermions in different layers as if they
were oppositely charged. Thus, while the coherent propagation of a
single pseudo-fermion is strongly inhibited by the random, path
dependent Aharonov-Bohm phase coming from fluctuations in $a^{(-)}_1$,
a {\it pair} of pseudo-fermions, one from each layer, {\it can}
propagate coherently through these fluctuations, because the
Aharonov-Bohm phase seen by one pseudo-fermion exactly cancels that
seen by the second. The physics here is remarkably similar to the
problem of holes constrained to hop on the same sublattice of a
quantum disordered antiferromagnet; a problem which can be mapped onto
an effective field theory in which holes on different sublattices
interact via a fictitious gauge field as if they were oppositely
charged \cite{lee-wen}.  In this problem, a pair condensate is
expected to appear \cite{singularpairing}, and it is interesting to
note that if a similar pseudo-fermion pair condensate,
$\langle\psi_1^*\psi_2^*\rangle \neq 0$, were to appear in the
$\nu=1/2$ double-layer system, it would correspond to a FQHE for the
original electrons, in the same way that the $\nu=1/3$ single-layer
FQHE can be understood in terms of the condensation of pseudo-bosons
\cite{zhang}.  Such a pairing scenario is similar to that considered
by Greiter {\it et al.} \cite{greiter} for the spin-polarized
$\nu=1/2$ state, where pairing of like spin electrons in the $p$-wave
channel was argued to lead to a single-layer spin-polarized $\nu=1/2$
FQHE.  However, in the double-layer system it is unlikely that the
transition from the (4,0) `metallic' phase to the (3,1) `condensed'
phase can be understood simply in terms of the effective pairing
interaction (\ref{veff3}), particularly because this interaction
changes when the system becomes superconducting.  In fact, the
experimental evidence for a compressible phase when the layer spacing
is large enough indicates that, in general, this pairing instability
does not occur.  Nevertheless, the above arguments suggest that there
should be pairing fluctuations present in the compressible phase of a
$\nu=1/2$ double-layer system; fluctuations which grow stronger with
decreasing $d$, and which, perhaps, play some role in the eventual
instability of the (4,0) phase to the (3,1) phase.

A rough measure of the relative importance of fluctuations in
$a^{(+)}_1$ and $a^{(-)}_1$ can be found by calculating the scattering
rates due to these two types of fluctuations for a pseudo-fermion with
initial energy $\epsilon_{\bf k}$.  Following similar analyses in the
literature \cite{halperin,lee-nagaosa}, we use Fermi's golden rule to
obtain
\begin{equation}
{1\over\tau^{+}_k} \simeq {k_f \varepsilon \phi^2\over{8m_be^2}}
\epsilon_{\bf k}
\end{equation}
for scattering from the symmetric fluctuations, and
\begin{equation}
{1\over\tau^{-}_k} \simeq {3\sqrt{3}\over{32\pi}}
\left({8\pi E_f\over{m_b^2\chi^2}}\right)^{1/3} \epsilon_{\bf k}^{2/3}
\end{equation}
for scattering from the antisymmetric fluctuations. Here $E_f =
k_f^2/2m_b = \omega_c/\phi$ is the mean-field Fermi energy.  Because
$a_1^{(-)}$ leads to more singular scattering than $a_1^{(+)}$, there
is a region around the Fermi surface, $|k^2/2m_b - \mu_f| \alt E_c$,
for which scattering from $a^{(-)}_1$ is dominant, where
\begin{equation}
{E_c\over{E_f}} \simeq {(3\sqrt{3})^3\over{4}} {1\over{\phi^2
(1+\phi^2/6 +d/a_0^*)^2 (k_f a^*_0)^3}}.\label{energy}
\end{equation}
Here all dependence on the band mass has been absorbed into $a^*_0
\equiv \varepsilon/e^2m_b$, the effective Bohr radius, ($a^*_0 \simeq$
82 \AA\ for GaAs).  As noted in Ref.~\cite{halperin}, in the extreme
quantum limit, where $a_0^* \gg l_0$, the kinetic energy is completely
quenched by the applied field and the band mass should not appear in
any physically relevant low-energy quantities.  In this limit we
expect that corrections beyond the random-phase approximation will
effectively renormalize $a_0^*$ to $l_0$ in (\ref{energy}).  However,
for the experiments discussed in \cite{eisenstein}, $l_0 \sim a^*_0$,
and the system is {\it not} in the extreme quantum limit.  Accordingly
in what follows we have used (\ref{energy}) without modification.
Table I summarizes the parameters $k_f a^*_0$, $d/a^*_0$, $E_f$ and
$E_c/E_f$ which characterize the four samples discussed in
Ref.~\cite{eisenstein}.  For samples A, B, and C, which exhibit the
FQHE, and which are, presumably, in the (3,1) phase, $E_c$ becomes
smaller as the observed FQHE weakens.  And for sample D, which does
not show the FQHE at all, $E_c$ has the smallest value of all four
samples.  This is consistent with the hypothesis that the out-of-phase
fluctuations are related to the instability of the (4,0) phase to the
(3,1) phase.

To conclude, the compressible phase of a double-layer electron system
in a transverse magnetic field with total Landau-level filling factor
$\nu = 1/2$ has been studied.  Following Halperin, Lee, and Read
\cite{halperin}, and Kalmeyer and Zhang \cite{kalmeyer}, the system
was transformed into a mathematically equivalent system of
pseudo-fermions in zero average magnetic field interacting via a
Chern-Simons gauge field, as well as the interlayer and intralayer
Coulomb repulsion.  The two layers decouple naturally in the
$d\rightarrow\infty$ limit, but for finite $d$ the interlayer Coulomb
repulsion gives rise to a new, low-lying diffusive mode.  This new
mode leads both to more singular low momentum scattering than occurs
in the single layer case, and to an attractive pairing interaction
between pseudo-fermions in different layers.  The appearance of this
attractive interaction, which grows stronger with decreasing $d$, may
be related to the experimentally observed instability of the
compressible (4,0) phase to an incompressible FQHE state.

I would like to acknowledge useful discussions with D. Khveshchenko,
D.H. Lee, L. Lilly, M. Reizer, J.R. Schrieffer, F.C. Zhang, and S.C.
Zhang.  This work was supported by NSF Grant No. DMR-91-14553 and by
the National High Magnetic Field Laboratory at Florida State
University.

\begin{table}
\caption{Sample parameters from Ref.~\protect\cite{eisenstein}.
$a^*_0$ is the effective Bohr radius, $d$ is the layer spacing, $k_f$
and $E_f$ are the mean-field Fermi wave vector and energy, and $E_c$
is the energy scale defined in (\protect\ref{energy}).}
\label{table}
\begin{tabular}{ccccccc}
 & & & $E_f$ & $E_c$ & &{\rm Strength} \\ Sample&$d/a^*_0$&
$k_fa^*_0$&(meV)&(meV)& $E_c/E_f$ & of $\nu$=1/2 \\ \tableline
A & 2.6 & 0.66 & 3.4 & 0.67 & 0.19 & {\rm Strongest} \\
B & 2.6 & 0.75 & 4.5 & 0.60 & 0.13 & {\rm Strong}    \\
C & 2.6 & 0.80 & 5.1 & 0.56 & 0.11 & {\rm Weak}      \\
D & 3.4 & 0.75 & 4.5 & 0.47 & 0.10 & {\rm Absent}    \\
\end{tabular}

\end{table}

\end{document}